# Active Architecture for Pervasive Contextual Services

Graham Kirby, Alan Dearle, Ron Morrison
*School of Computer Science, University of St Andrews,*
*North Haugh, St Andrews, Fife KY16 9SS, Scotland*
{graham, al, ron}@dcs.st-and.ac.uk

Mark Dunlop, Richard Connor, Paddy Nixon
*Department of Computer and Information Sciences, University of Strathclyde,*
*26 Richmond Street, Glasgow G1 1XH, Scotland*
{Mark.Dunlop, Richard.Connor, Paddy.Nixon}@cis.strath.ac.uk

## Abstract

Pervasive services may be defined as services that are available *to any client (anytime, anywhere)*. Here we focus on the software and network infrastructure required to support pervasive contextual services operating over a wide area. One of the key requirements is a matching service capable of assimilating and filtering information from various sources and determining matches relevant to those services. We consider some of the challenges in engineering a globally distributed matching service that is scalable, manageable, and able to evolve incrementally as usage patterns, data formats, services, network topologies and deployment technologies change. We outline an approach based on the use of a peer-to-peer architecture to distribute user events and data, and to support the deployment and evolution of the infrastructure itself.

## 1 Introduction

### 1.1 Requirements

Pervasive services may be defined as services that are available *to any client (anytime, anywhere)* [1]. From this definition the global nature of pervasive services is implicit. Contextual services may be defined as any services that take account of the user's current context; many research projects are attempting to define context, e.g. [2, 3]. One useful definition is given in [4] as:

*"We call the environment in which a computation takes place its context. We define context in the broadest possible sense. A context may either refer to aspects of the physical world or to conditions and activities in the virtual world."*

Examples of contextual information include: time, location (both coordinate and logical location), current personal role, physical attributes of the environment (e.g. temperature, ambient light), etc.

Most research on contextual services concentrates on local environments. Here we focus on the software and network infrastructure required to support pervasive contextual services operating over a wide area.

One of the key requirements to the provision of globally pervasive contextual services is a **matching service** capable of assimilating and filtering information from various sources and determining relevant matches. Information sources include contextual information from the environment, personal preferences, user history etc, relatively static information such as spatial data from GIS, and more general information published on intranets and the internet. A matching service can be considered to be an entity that, triggered by the reception of events from multiple sources, synthesises a stream of new events. Typically, the output events will be higher-level (more semantically meaningful) than the input events. In addition to the input event streams, the matching service will operate over a global knowledge base comprising elements such as GIS, web-based systems, databases, semi-structured data, etc.

Each output event describes a correlation of input events and facts that is relevant to a contextual service. For example, these items might be correlated:

- user Bob likes ice cream, but only when the weather is hot and when he has spare time to eat it
- it is 20ºC in South Street is at 16.30 on 25/6/2003
- Bob is on holiday from 20/6/2003 to 27/6/2003
- Bob is Scottish
- Bob is in North Street at 16.45 on 25/6/2003
- Bob is on foot on 25/6/2003
- Janetta's in Market Street sells ice cream, and is open between 9.00 and 17.00
- Bob knows Anna
- Anna is at coordinate 56.3397, -2.80753 at 16.15 on 25/6/2003

If, within the time interval 16.45-16.50, all these items could be correlated, a pervasive contextual service could suggest to both Bob and Anna via some appropriate user interface mechanism that they might wish to meet for an ice cream at Janetta's at 16.55. This correlation requires the detection of spatial, temporal and logical relationships; it can be inferred from the set of items that both Bob and Anna are probably close enough to Janetta's to get there before it closes. Similarly, it can be inferred that Bob would probably like an ice cream given that he is Scottish and therefore regards 20º as hot. It is relatively straightforward to make these inferences if the small set of items is known; the major difficulty is in extracting the correlated set in the first place, from the huge number of items available.

Further difficulties arise: although the example above was relatively localised in space, in other scenarios the items to be matched might be globally distributed—e.g. Bob, currently in Australia, walks past a restaurant previously recommended by Anna: her opinion of the restaurant should delivered to Bob if it is dinner time and has no plans for dinner, or if he is staying a few more days in the area. Furthermore, matching for a range of different services must take place simultaneously, for the entire population of participating users, as indicated in Figure 1. The key point is that delivery of pervasive contextual services requires the continuous processing of a very high volume of globally distributed items of information, distilling them down into a relatively small volume of meaningful events.

## 1.2 Refining the Requirements

If we accept these requirements for a global contextual matching service, the question arises as to whether such an entity can be engineered, and if so, how. The service cannot be provided on a single machine or in a single location; therefore, in addition to the items being matched, the matching computation must itself be distributed. We can refine the requirements as being a distributed matching engine, fed from a distributed event system, and a distributed knowledge base.

Clearly the construction of such an entity is itself a grand challenge. There are many individual engineering challenges: information must be assimilated from numerous sources, requiring ontologies for describing it and tools for extracting it; databases are required for storing the information; the construction of the distributed matching engine in itself is extremely complex; and finally, the system must be engineered to operate with various hand-held and other mobile technologies.

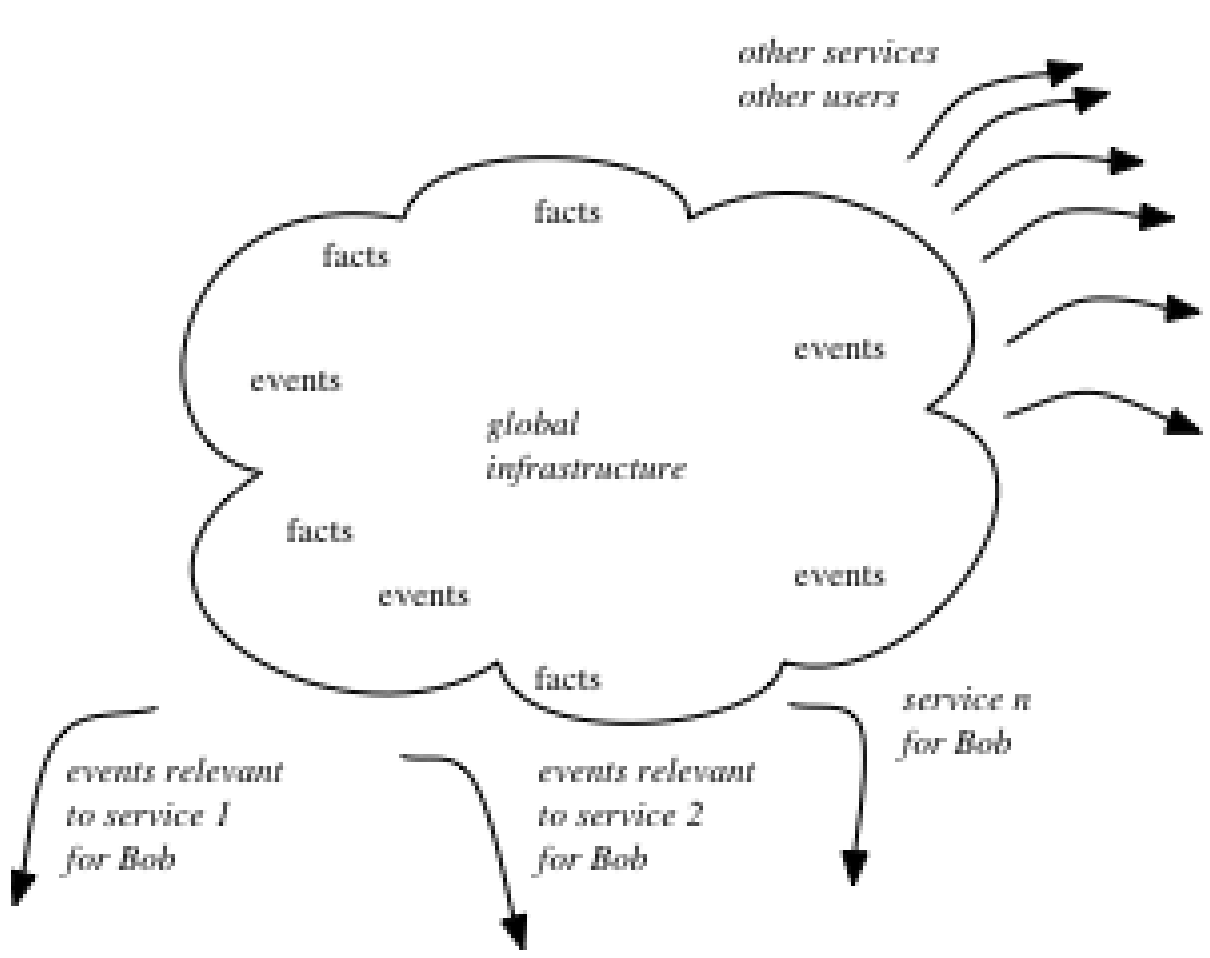

Figure 1: A global matching service.

As with any large-scale distributed system, the development of a wide area infrastructure for pervasive contextual services poses a number of novel research challenges in terms of evolution, management and scalability; these problems are not significant in the local environments in which most contextual services operate. The distributed matching engine must have the ability to evolve with, and adapt to, changing usage patterns, data formats, services, network topologies and deployment technologies. Given its scale and non-centralised nature, it will be impossible to shut it down and restart it for maintenance. Clearly then, the system must be capable of incrementally changing its configuration. This assumes, of course, that data and systems can be organised appropriately to make use of available resources without imposing complexity on the user. The problems of complexity in current systems, even on the small scale, are identified by Bolosky:

*"Users are subjected to random performance and service disruptions. Replacing or upgrading a personal computer, workstation, or server is very difficult. Even a moderate size computer network requires significant expertise to configure and maintain."*

In the distributed matching engine data must flow around the system in response to: changes in users' location and behaviour; changes in the access patterns of processes; changes in the physical resources allocated to the system; or changes in the topology of the physical infrastructure. It is essential for the underlying policies to evolve in response to such changes, but the complexity is such that it is infeasible for this to be controlled by human users or administrators. The system must therefore manage such changes automatically. In order to achieve scalability, the system needs to be capable of making best use of the resources that are available at any time. Ideally, the system would be able to intercept and utilise network, storage and compute power as it became available in any location.

The matching engine is required to match events produced in real-time by real-world interactions. The matching process must be capable of processing the event stream sufficiently quickly to produce contextual information that is pertinent to users within an appropriate time frame. In order to do this matching, both the events and the knowledge base must be delivered to the locations at which the matching computation occurs. Clearly this imposes requirements on the event delivery system and the caching and replication policies. In particular, if the data is not available with sufficiently low latency, matching cannot occur at the necessary rate. Thus the infrastructure must facilitate the appropriate placement of **data** and **computation** at all times—on servers, end user devices, and within the network.

## 2 Challenges

The challenges in meeting the requirements outlined above include:

- the provision of a generic global event service capable of: delivering events to appropriate locations at appropriate times; and delivering multiple event types including those unknown at initial deployment;
- the engineering of the matching computation, including: the partitioning of the computation; deployment of the components at appropriate locations at appropriate times; incorporation and dropping of storage, computational and network resources as they become available/unavailable; data caching and replication policies to ensure that the matching process can occur sufficiently fast and with sufficient reliability; ontologies and type systems for describing data, knowledge and events—and flexible mechanisms for binding to these; adaptation to changing patterns of use by individuals and populations;
- the ability to incorporate new devices and technologies incrementally as they become available, given that it is not possible to halt and restart the system;
- the provision of infrastructure upon which new services may be implemented;
- programming abstractions with which to describe the services and how they should be deployed.

A number of other issues are crucial to a practical deployment, but not specifically addressed in this paper:

- security mechanisms to ensure that data and computations may be accessed and placed only in appropriate ways;
- privacy policies that allow users to control the degree to which sensitive data is stored in the global system and used in various ways;
- definition of user centred quality of service, annoyance, security and privacy requirements.

## 3 State of the Art

Many projects attempt to address the needs of an event distribution service for pervasive systems. Some have been deployed and work well in a relatively localised area. For example, Elvin [5] supports a content based addressing mechanism above a publish/subscribe network. The system is based on subscription to events supported by a subscription language; it uses a client-server architecture, limiting its scalability.

Siena [6] addresses scalability directly and aims to provide a wide area event notification system. Event producers advertise the events that they generate and event sinks subscribe to notifications of interest. Events are represented as 3-tuples of a name, type and value. Siena does not support mobility directly, however, Mobikit [7] extends Siena by providing a mobility service that supports both user mobility and code mobility. The system provides static proxies for mobile entities, which subscribe on behalf of the mobile entity when the mobile entity is disconnected from the pub/sub system.

The need to push and deploy code in the network is also being addressed by the network communities and active mobile agents. The Active Network community supports the deployment of *capsules* [8] containing both code and data. For example, the ANTS system supports a Java-based toolkit for experimenting with active networks in which code capsules serve as the minimum unit of deployment and are used as a measure of resource utilisation.

We are currently constructing the Cingal system, which provides code push technology permitting *bundles* of code and data wrapped in XML packets to be deployed and run on a *thin server*. On arrival at a thin server, and subject to verification and security checks, the code may be executed within a security domain. Each thin server provides the necessary infrastructure for code deployment, authentication of bundles, a capability-based protection system and an object store.

A plethora of Java based active mobile agent systems exists ranging from commercial products such as Voyager and Pathwalker to research projects such as Mole [9], HIVE [10] and ARA [11]. These cover the entire code migration spectrum from strong (in the case of ARA) to weak schemes (the rest). An agent is a process that migrates among the nodes within a network to perform its task, operating on behalf of a user. Every visited node must support an infrastructure to deal with the management of system services and resources. This is usually done via fully trusted non-mobile agents.

With *active pipes* [12], the capabilities of nodes and groups of nodes are described in terms of static and dynamic properties. Processing and transmission requirements are expressed as a sequence of functions to be performed on a data stream. These are mapped onto nodes by placing constraints on the processing elements such as: IP-range, network services supported, link-load etc. For each processing step, constraints define a subset of nodes qualified to execute a given function.

Constructing inferences from matched events is a critical aspect of any context infrastructure, providing the means by which services are targeted to the individual and situation. Core to this is the ability to match events from the real world, correlate against existing knowledge, and identify a meaningful action. The issue of describing and matching has been investigated in various domains. Proposed solutions fall into three categories: text based, lexical descriptor based and specification based. Text based solutions use the textual representation as an implicit description of behaviour, while employing arbitrarily complex string matching expressions. Although these have low maintenance cost and are easy to introduce, a textual representation does not guarantee sufficient information for the classification and in fact could be misleading. Lexical descriptor based solutions use key phrases, which are constructed from a predefined vocabulary provided by subject experts, to describe the target. This can be extended to describe a number of different aspects of the situation, leading to the technique of multi-faceted classification [13]. The use of key phrases makes the method sounder and more complete. The construction of the predefined vocabulary, or ontology, is a non-trivial task and there is also ambiguity associated with the type of semantics (computational or application) that the vocabulary should describe. Finally, specification based solutions use a specification language, whose semantics define the classification and retrieval scheme [14].

The logic programming, constraint and context-aware retrieval (CAR) communities are developing technologies that address the needs of a contextual matching service. In logic programming, declarative programming languages such as Prolog can be used to express matches using first order logic. Distributed Prolog systems have been constructed, notably SICStus Prolog which supports distributed unification. The constraint community have produced constraint solvers that search for solutions using forward checking and constraint logic programming. Research on the Distributed Constraint Satisfaction Problem extends this by distributing the variables to be solved amongst agents.

The Rome system [15] supports the concept of a context *trigger*, consisting of a condition and an action. It allows decentralized evaluation of triggers by embedding triggers in end devices. This does not, however, allow context sharing and requires the end device to have the capability to sense and process all of the necessary raw contextual information, which may not be efficiently achieved, especially for a complicated trigger and a simple device.

XML is established as the *de facto* standard for information interchange; it is therefore reasonable to assume that both events and knowledge will be stored in an XML format. There is therefore a requirement for programs to be able to bind to XML data. Most programming interfaces incorporate a tree-structured view of XML as an abstraction for the programmer. This is a good abstraction if the data content is inherently irregular, however for regular data it leaves much to be desired in terms of both application modelling and runtime efficiency. A better approach is to present the programmer with an abstraction corresponding to a more traditional data model. There is a further option of how to achieve this; there are two possibilities:

- type generation, where a programming language type is obtained by analysis of either the data itself or a metadata description of it [16, 17], and
- type projection, where the type is taken from the program context and matched against the data [18, 19]

Our interest here is in the second strategy, which has various advantages. Crucially in this context, these include the ability to handle partial data model specifications. This is key in the case where the overall structure of the data is not tightly specified, yet it contains structured 'islands' whose structure is known *a priori*. This is a key requirement in this context, where there is inherently a lack of pre-imposed global standardisation and rapidly evolving data modelling requirements.

The PSI project [1] directly addresses infrastructure for pervasive services. It proposes a 3-layer architecture comprising front end clients such as PDAs, embedded devices etc.; infrastructure servers and backend servers. The project's thesis is that small front end devices can borrow resources from back end servers. This requires that applications be split into components and distributed appropriately. Currently, the partitioning of applications appears to be being done manually. However other projects such as RAFDA [20] and Coign [21] are attempting to do this automatically. For example, the RAFDA project is investigating partitioning Java applications into components capable of being executed on different virtual machines.

Recently a new generation of storage architectures has emerged based on Peer-To-Peer (P2P) technologies. These include OceanStore [22], Mnemosyne [23], PAST [24], Pastry [25], FreeHaven [26] and Freenet [27], and all provide some degree of abstraction over the location of data and utilise storage available in a network of peers. Most of the above are based on a deterministic routing algorithm by Plaxton [28], which permits the discovery of documents stored in a wide area network such as the Internet. Some systems, such as [27], rely exclusively on non-deterministic algorithms. This mean that data cannot always be found, rendering them unsuitable as a base technology for this work. All the P2P architectures cited use hashing algorithms to assign each document with a globally unique identifier (GUID). Typically this is either derived purely from document content using secure hashes, or from a hash of keywords, filename and the public key of the creator.

Many of the cited systems store multiple replicated copies of data to provide resilience and to increase the probability that copies of a data item will be stored close (in terms of latency) to its readers. OceanStore in particular supports algorithms for distributing data globally. The schemes for storing replicated copies of data vary from simple block copying to erasure-codes which permit data to be reconstituted from a subset of the servers on which it is stored. The more sophisticated P2P systems support *promiscuous caching* where data is free to be cached anywhere at any time. This does not affect the correctness of the system nor prohibit other users from manipulating that data, and is crucial to the performance of the system if the fetching of remote data at every access is to be avoided.

## 4 Our Position

We now outline our approach to addressing the challenges presented. Broadly, this is to develop an active network whose topology and memberships can change dynamically, defined over the embedded sensors, mobile devices, servers and the networks that link them. Each node—mobile device, server or network component—stores information, computes over it, and communicates with other nodes. Dynamicity is achieved by pushing code and data onto the network to define data channels, processing rules and routing rules.

### 4.1 Generic Global Event Service

We believe that P2P architectures are highly suited to the provision of pervasive services. In particular, a P2P architecture may be used to distribute both low-level sensor-derived events, and high-level synthesised events produced by the contextual matching engine. We propose that a general-purpose system such as Siena would be ideal for this purpose. It has enough expressibility in its publish/subscribe language and shows evidence of being globally scalable.

### 4.2 Partitioning of Matching Computation

Our approach is to implement a distributed contextual matching engine as XML pipelines [29], with XML events flowing between pipeline components, both intra-node and inter-node. This scheme is designed to abstract over any particular technology, and to make pipeline components independent of each other.

XML event buses allow incoming events to be delivered to multiple downstream components, which may reside on the same node or on remote nodes. Each pipeline provides a web service interface *put(event)*, enabling remote pipeline components to push events into it. Events may also arise from local devices and sensors such as GPS and GSM devices, RFID tag readers, weather sensors, etc. Each hardware device has a wrapper component that makes it usable as a pipeline component. Other components perform filtering (e.g. transmitting user-location events only when the distance moved exceeds a certain threshold), buffering, communication with other pipelines, and so on.

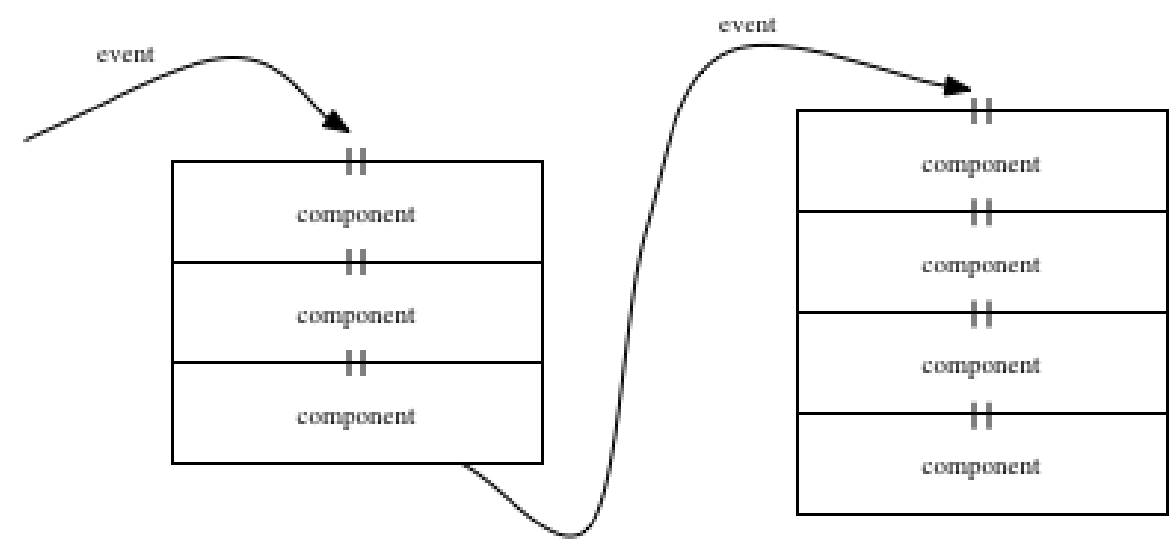

Figure 2: Distributed XML pipelines.

Web service interfaces permit interoperability between heterogeneous platforms and languages and define how pipeline components transmit and receive events. The interconnection topology is orthogonal to the service definition and its deployment, and may be provided via a P2P system or a traditional client-server architecture. Figure 2 shows an example of a pipeline distributed over two nodes.

### 4.3 Component Deployment Mechanisms

Given the approach to partitioning the matching computation using distributed event pipelines, we require mechanisms to deploy and evolve the pipeline components. The mapping of pipeline component types to physical nodes is likely to be extremely complex, and we consider it infeasible to fully establish this mapping statically, before deployment. Instead, we propose to break the deployment problem down into:

- initial deployment of a pipeline deployment infrastructure onto all participating nodes, followed by:
- ongoing deployment and redeployment of individual pipeline components onto those nodes, using the deployment infrastructure already loaded.

This allows flexibility in establishing an initial deployment, and also for subsequent evolution of the pipelines in response to changes in resource availability, user demand, access patterns etc. Rather than trying to address the need for evolution in a post-hoc manner, this architecture is designed around the ability to evolve from its inception. Indeed, we view deployment of the architecture as being an evolutionary step itself.

The pipeline deployment infrastructure will itself be deployed on participating nodes using any conventional software installation approach. The infrastructure will be based on code push technology developed in our Cingal project. We propose to exploit this by constructing the pipeline components as code bundles that may be deployed onto Cingal thin servers. This will provide flexibility both in initial deployment and in later incremental evolution of the components and their topology. Figure 3 shows an example of a thin server node, containing both the deployment infrastructure and a pipeline assembled from components sent to it in code bundles.

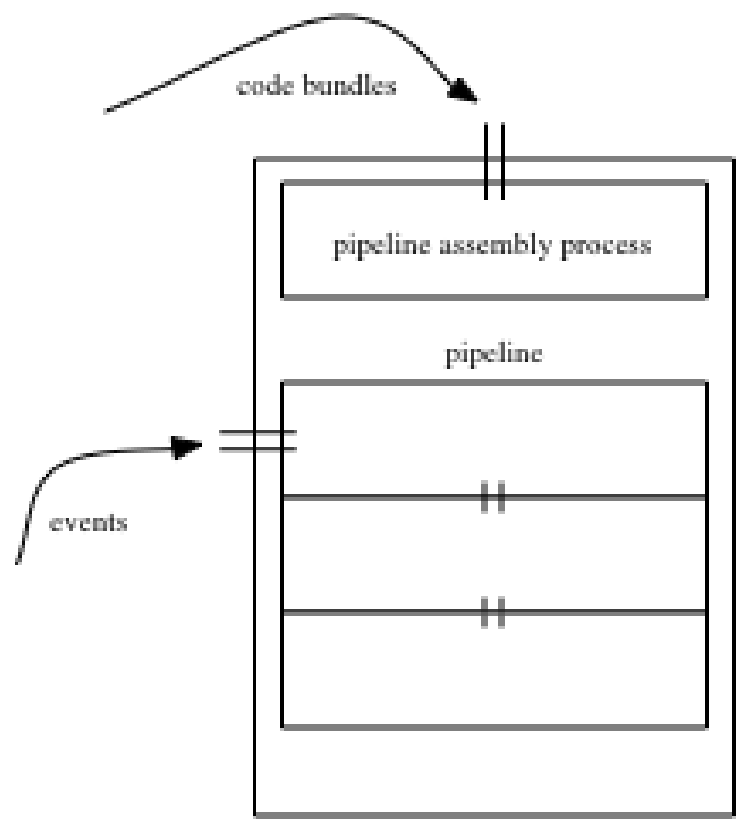

Figure 3: Pipeline deployment infrastructure.

## 4.4 Component Deployment Policies

The mechanisms described in the previous section will support flexible policies for mapping the pipelines comprising the contextual matching engine across the network. The need for flexibility arises because the architecture must be able to evolve, as already discussed, and also because initially specified policies will often require refinement in the light of evaluation following deployment.

Implementers and administrators will need to be able to express deployment policies at a suitable level of abstraction. Given that we wish to take advantage of new computational, storage and network resources as they become available dynamically—and, conversely, to adapt gracefully when they disappear—these policies should not be expressed in terms of specific individual physical resources. We adopt the active pipe approach [12], in which policies take the form of constraints over the placement of processing steps. For example, a constraint might specify that at least 5 pipeline components providing a data replication service must be deployed in parallel within a given geographical region.

The set of constraints used in active pipes will be extended to include higher-level concepts such as dynamic caching and replication policies, which are crucial to the provision of pervasive contextual services. All constraints will feed into an evolution engine, itself a distributed computation, that will dynamically evolve the contextual matching engine by manipulating the pipelines as described in the previous section. As events arise that cause a given constraint to be violated (such as the sudden unavailability of a particular node), it is the role of the monitoring engine to make appropriate adjustments to satisfy the constraint again.

This approach requires knowledge of available resources to be propagated dynamically to the evolution engine, which is itself decentralised. As with user events, we propose to use a P2P system to propagate events describing changes in resource availability. Nodes will advertise their resource availability, physical and logical connectivity, geographic location etc. via *publish* events on a P2P system. The events may be subscribed to by the evolution engine, which can then elect to make use of these resources by deploying new pipeline components onto them. Nodes may disappear from the network either gracefully, in which case they will publish events warning of their imminent withdrawal, or without warning, in which case the loss may eventually be detected by other monitoring components, which will publish events on their behalf.

## 4.5 Caching and Replication of Data

As stated earlier, for the matching service to operate with sufficiently low latency, appropriate caching and replication policies and mechanisms must be available. The use of *promiscuous caching* described earlier combined with a global storage architecture such as one of the schemes based on Plaxton routing appears an ideal combination for the global matching engine.

## 4.6 Adapting To Changing Use Patterns

Efficient operation of the contextual matching engine will require adaptation to changing patterns of use by individuals and populations. Data will flow around the system in response to changes in users' location and behaviour, and changes in their patterns of access to information. The system might observe diurnal patterns in data access at a microscopic level associated with a particular user, or at a macroscopic level on a global scale. In response to these observations the system would modify the constraint set to optimise the caching and replication of data as is appropriate.

In order to achieve this, the system must monitor: the physical infrastructure; behaviour of users including their physical location(s) and patterns of access to data; and the internal state of the system itself. The last point is important since, for example, a rule might create 5 copies of some data for resilience, but over time some of these might become unavailable—in which case further copies should be made. An obvious analogy is with RAID systems, which self-heal in response to individual component failure.

Data placement monitors will observe meta-data arising from distributed probes and gauges. Periodically they will initiate data replication, the details of when and where depending on the placement policies currently in operation. Various policies may be designed to pursue different goals. A *latency-reduction* policy might, for example, seek to replicate progressively more of a user's personal data at storage units geographically close to the user's current location, the longer that the user remained at that location. A *backup* policy might seek to replicate data on a geographically remote storage unit as soon as possible after it was created.

## 4.7 Adapting To Changes In Technology

Given the issues of large-scale deployment already discussed, it will be essential that improvements in technology can be incorporated without requiring the entire system to be re-deployed. This applies both to external software services and to hardware. Our approach is to use standardised and open interfaces and data formats wherever possible—thus XML-encoded events, web service interfaces for pushing events and new code bundles, etc. This also supports software and hardware heterogeneity since multiple implementations of the interfaces may inter-operate.

## 4.8 Service Infrastructure

We expect that many different pervasive contextual services will be possible, and that the speed with which new services can be introduced will be commercially significant. It will thus be important to provide a common software infrastructure upon which new services can be implemented. This will factor out software that is relevant to multiple services for reuse. The nature of the operations provided by the infrastructure is an open issue; our initial set of operations includes APIs to the event pipelines, to the code bundle deployment mechanism, and to the distributed contextual matching engine.

## 4.9 Programming Abstractions

Related to the previous section, it will be important also to develop suitable abstractions over the software infrastructure, in order to be able to express the essentials of a new pervasive contextual service without being overwhelmed by the details. Where appropriate, the developer should be able to largely ignore issues of precise network topology and individual node failure, and concentrate on the fundamental aspects of the new service—what information should be delivered to the user, in what form, and in which context. Our approach here is to develop declarative notations to describe the placement of computation and data, allowing the developer to write constraints that feed into the deployment evolution engine.

# 5 Conclusions

The overall system architecture consists of several P2P systems overlaid on each other in order to implement and support the global matching engine. An event system delivers events from users and sensors. These include user location events, temperature readings etc. Machines advertise their arrival and departure from the matching engine. Some events may be generated by machines that are monitoring the machines available to the engine. Both classes of events are supported by a Siena-like P2P system. The caching and replication of data is handled by a Plaxton based storage architecture supported by promiscuous caching mechanisms.

When new computational or storage resources are detected by the matching engine, computations are pushed onto them as code bundles using technology developed in the Cingal project. Once installed, these computations can offer additional computational resources for the matching engine (*matchlets*) or provide storage capacity for the storage architecture (*storelets*).

Matchlets are structured as pipeline code that accepts events from the event distribution mechanism and performs matching on them. Each matchlet writes its results onto the event bus. Thus the primary API offered by the host to matchlets is an event delivery source and an event sink. Matchlets use type projection mechanisms for binding to the XML data contained within the events.

As the system evolves, new event types will be introduced. In order to deal with unknown events, a mechanism is needed within the event distribution mechanism for routing unknown event types to *discovery matchlets*. These look for code capable of matching these new events in the storage architecture and deploy this code onto the network. Of course, the deployment of any bundles needs to be consistent with the constraint rules associated with them in order to ensure semantic consistency. Such constraints must be enforced dynamically.

# 6 Acknowledgements

This work was supported by the FP5 Gloss project IST-2000-26070, with partners at Trinity College Dublin and Université Joseph Fourier, and by EPSRC grants GR/M78403/GR/M76225, *Supporting Internet Computation in Arbitrary Geographical Locations*, and GR/R45154, *Bulk Storage of XML Documents*.